# Structure and stability of 7:3 rare earth oxide-phosphates:

## a combined ab initio and experimental study

Ligen Wang[a], Konrad Burkmann[b], Sergey V. Ushakov[b], Edric X. Wang[a,c], Jared Matteucci[b], Mara Scheuermann[d], Erik Melnitschuk[d], Robert Glaum[d], Hongwu Xu[b], Elizabeth J. Opila[e], Alexandra Navrotsky[a,b], Qi-Jun Hong[a,*]

[a] Materials Science and Engineering, Arizona State University, Tempe, AZ 85287, USA

[b] Navrotsky-Eyring Center for Materials of the Universe, School of Molecular Sciences, Arizona State University, Tempe, AZ 85287, USA

[c] Department of Materials Science and Engineering, University of Illinois at Urbana-Champaign, Urbana, IL 61801, USA

[d] Department of Inorganic Chemistry, Rheinische Friedrich-Wilhelms-Universität Bonn, Gerhard-Domagk-Straße 1, 53121 Bonn, Germany

[e] Department of Materials Science and Engineering, University of Virginia, Charlottesville, VA 22903, USA

## Abstract

Rare earth oxide-phosphates (REOPs) form a largely unexplored family of refractory lanthanides and yttrium compounds with general formula $RE_xO_y(PO_4)_z$. They are of interest for applications ranging from thermal barrier coatings to catalysts and magnetic materials. At least four REOPs phases were experimentally identified with RE/P ratios from 7:3 to 6:1, however the structures were solved only for 3:1 phases ($RE_3O_3(PO_4)$). In this work we report the structure for the 7:3 phases ($RE_7O_6(PO_4)_3$) derived by ab initio analysis of models based on previously reported oxide-vanadate analogues. The most stable structures for all 7:3 REOPs were found to be isotypic, adopting monoclinic symmetry with space group $P2_1/c$. The structures were validated by comparison of their powder X-ray diffraction patterns to those of synthesized La, Pr, Nd, Sm, Eu, Gd and Tb 7:3 phases (Rietveld refinement for all except Tb). Ab initio analysis of thermodynamic stability showed that all 7:3 REOPs are unstable at 0 K toward decomposition to $REPO_4$ and $RE_3PO_7$ or $RE_2O_3$. The entropy contribution stabilizes $RE_7O_6(PO_4)_3$ phases for light rare earth elements above 1000 K, however, starting with Dy, computationally predicted stabilization temperature is higher than estimated melting points of $RE_7O_6(PO_4)_3$, which is consistent with observed synthesis pattern.

***Keywords:*** *Density functional theory calculations, Rare earth oxide-phosphates, Formation*

*energies, Thermodynamic stability, Vibrational entropy*

# 1. Introduction

Rare earth oxide-phosphates (REOPs) are compounds with general formula $RE_xO_y(PO_4)_z$, where (RE) refers to lanthanides and Y. This notation emphasizes that in REOPs, as in orthophosphates, phosphorus atoms are present in the structure as isolated $PO_4^{3-}$ tetrahedra, in contrast to the large family of polyphosphates in which they are linked by sharing oxygen atoms. RE/P ratio or reduced formula $RE_xP_yO_z$ is commonly used and we will follow these notations for brevity. For rare earth orthophosphates ($REPO_4$) the RE/P ratio is equal to 1, it is more than 1 for all REOPs, and less than 1 for all polyphosphates. REOPs were first discovered over half a century ago by melting of RE orthophosphates in a solar furnace [1]. Several stoichiometries were reported for REOPs of each rare earth element (Fig. 1) [1-8]. High melting temperatures, chemical stability, and structural diversity of REOPs make them promising candidates for design of new refractory coatings, catalysts, magnetic and luminescent materials. However, the advancement of REOP-based functional materials is currently limited by a lack of structural and thermodynamic data. To most REOPs monoclinic unit cells were assigned with large lattice parameters (e.g. 12 formula units in $RE_3PO_7$). Only the crystal structures of $Nd_3PO_7$ and $Eu_3PO_7$ have been experimentally determined using single-crystal X-ray diffraction [3,9], and only preliminary structural models have been proposed for several REOPs [2]. Recently, we reported ab initio structure optimization for all 3:1 REOPs, based on known structures, analyzed their stability with respect to $RE_2O_3$ and $REPO_4$, predicted and experimentally confirmed the existence of 3:1 and 7:3 cerium phases [7].

In this study, we performed a combined computational and experimental investigation of the crystal structure and phase stability of $RE_7P_3O_{18}$. The structures of two previously reported $La_7V_3O_{18}$ polymorphs [10] with vanadium replaced by phosphorus were used as starting models for ab initio analysis. Ab initio calculations revealed the β-$La_7V_3O_{18}$-type ($P2_1/c$) structure to be the more stable form for all $RE_7P_3O_{18}$ compounds.

Formation energies were calculated for all $RE_7P_3O_{18}$ compounds (where RE includes lanthanides and yttrium). The results indicate that at 0 K the $RE_7P_3O_{18}$ compounds are thermodynamically unstable toward decomposition into $REPO_4$ and $RE_3PO_7$ or $RE_2O_3$, lying within 55 meV/atom above the convex hull. However, $RE_7P_3O_{18}$ with RE = La–Tb are stabilized at high temperatures due to entropy contributions. The predicted structures for RE = La, Ce, Pr, Nd, Sm, Eu and Gd provided satisfactory fits when performing Rietveld refinement on the structures of synthesized $RE_7P_3O_{18}$ phases, thus rendering confidence in computational prediction of thermodynamic properties.

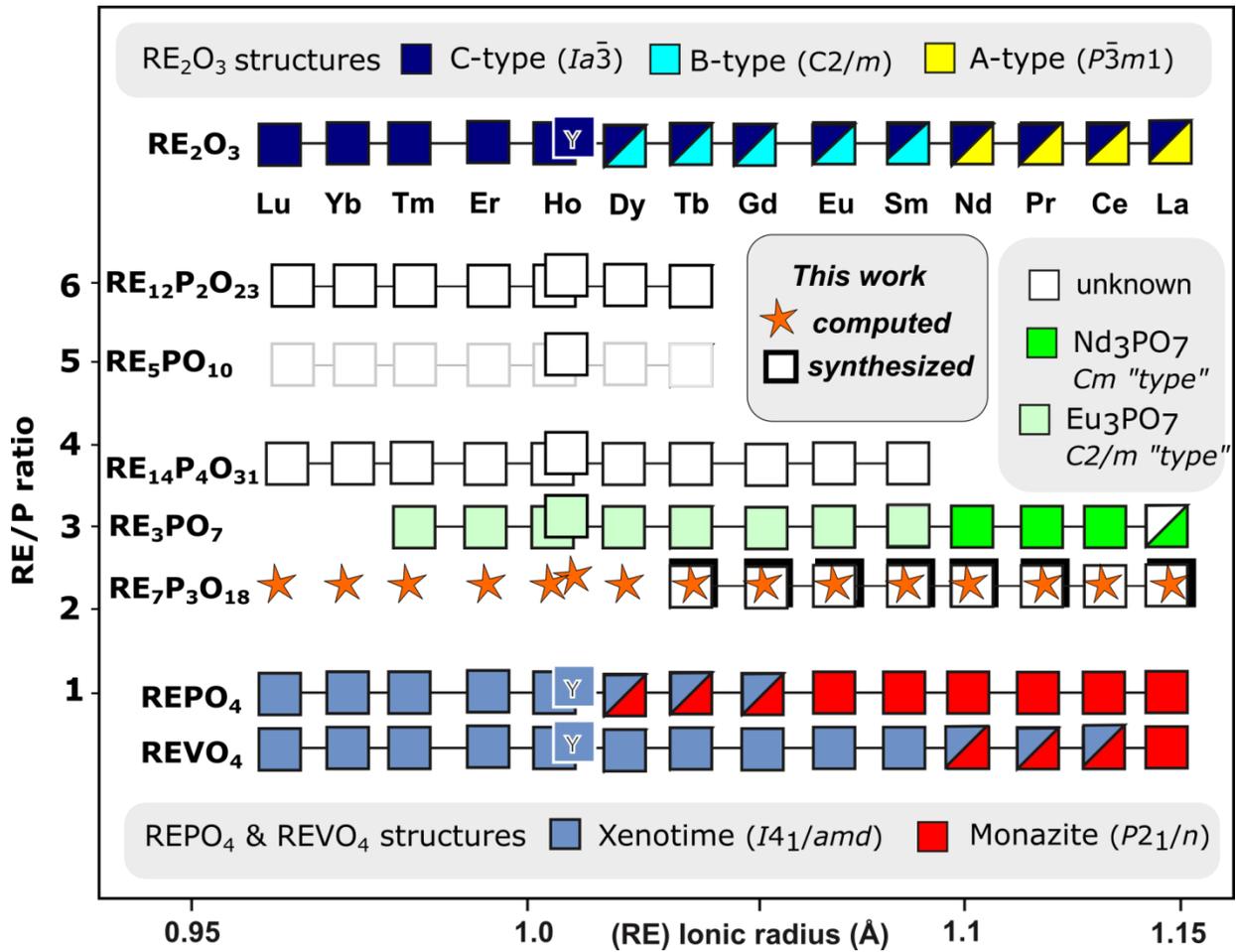

**Fig. 1.** Stoichiometries and structure types reported in REPO$_4$ - RE$_2$O$_3$ systems [1-6] and compounds computed and synthesized in this work. Ce$_7$P$_3$O$_{18}$ synthesis by laser melting of CePO$_4$ was reported earlier [7], and the X-ray diffraction data were analyzed in this work using DFT derived structure model. The occurrence of monazite and xenotime structures in REVO$_4$ is included for comparison, following Ref. [8].

## 2. Computational and experimental methods

### 2.1 Computational details

All calculations were performed within the framework of density functional theory (DFT) using the Vienna *Ab initio* Simulation Package (VASP) [11,12]. Projector augmented wave (PAW) pseudopotentials [13] were employed to model the interactions between ions and valence electrons. For the lanthanides Ce–Lu, pseudopotentials with a formal valence of +3 were used, in which the number of 4f electrons frozen in the core equals the total number of valence electrons minus the formal valency [14]. For example, samarium (Sm) has eight valence electrons—six 4f and two 6s electrons. Given its common trivalent state, five of the 4f electrons were treated as core electrons in the Sm_3 pseudopotential. For La, which has no occupied f states, the standard La pseudopotential was used. For yttrium (Y), the semicore 4s and 4p states were included in the valence using the Y_sv pseudopotential. The generalized gradient approximation (GGA) with the Perdew–Burke–Ernzerhof (PBE) exchange-correlation functional [15] was employed, as it offers a good balance between computational efficiency and accuracy. A plane-wave energy cutoff of 520 eV was used throughout. The Brillouin zone was sampled using k-point meshes [16] that were structurally similar to, but denser than, those used in the Materials Project database [17] for the $RE_2O_3$, $REPO_4$, and $RE_3PO_7$ phases. For the $RE_7P_3O_{18}$ phase, which is not included in the Materials Project database [17], we used an automatic k-point mesh with $R_k = 20$ [11,12]. Convergence of the formation enthalpy with respect to k-point density was carefully tested. Supercell parameters and atomic positions were optimized until the total energy converged within $10^{-5}$ eV and all atomic forces were minimized to below 0.02 eV/Å.

Phase stability can be assessed using the convex hull formalism [18], i.e. evaluating with respect to various possible phase-decomposition reactions. To assess the (in)stability of the $RE_7P_3O_{18}$ phase, the formation energies for forming $RE_7P_3O_{18}$ from $RE_3PO_7$ and $REPO_4$ or from $RE_2O_3$ and $REPO_4$ are defined below:

$$\Delta E_f(1) = E_{tot}(RE_7P_3O_{18}) - 2E_{tot}(RE_3PO_7) - E_{tot}(REPO_4), \qquad (1)$$

$$\Delta E_f(2) = E_{tot}(RE_7P_3O_{18}) - 2E_{tot}(RE_2O_3) - 3E_{tot}(REPO_4), \qquad (2)$$

where $E_{tot}(RE_7P_3O_{18})$, $E_{tot}(RE_3PO_7)$, $E_{tot}(RE_2O_3)$, and $E_{tot}(REPO_4)$ are the total energies for $RE_7P_3O_{18}$, $RE_3PO_7$, $RE_2O_3$, and $REPO_4$ in the bulk phases, respectively. For an unstable compound $\Delta E_f > 0$, whereas for a stable compound $\Delta E_f \leq 0$.

Lattice vibrational and configurational entropies significantly influence phase stability in materials [19]. Differences in vibrational entropy between phases arise from variations in atomic and bonding structures, including bond stiffness, volume changes, and atomic size mismatches. As demonstrated by Hong and Liu [20], a generalized methodology allows for the computation of entropy at various levels—configurational, vibrational, and electronic—from a single *ab initio* molecular dynamics (AIMD) trajectory in both solid and liquid phases. The quantum mechanical effects on quantum harmonic oscillation of phonons and the Fermi-Dirac distribution of electrons are simultaneously included in this approach. In this study, we applied their approach to calculate the free energies for various bulk phases, capturing all major entropy contributions in a unified framework. The formation free energies were calculated by substituting the free energies for the total energies in Eqs. (1) and (2). Consistent with the structural optimization and reaction energy calculations, the AIMD simulations were performed using the VASP package with the GGA-PBE exchange-correlation functional.

**2.2 Experimental details**

The $RE_7P_3O_{18}$ samples were synthesized using the reaction between diammonium hydrogen phosphate and rare earth sesquioxides. $RE_2O_3$ were obtained from Sigma Aldrich (Gd), Alfa Aesar (La, Nd, Sm, Eu, Tb) with metal-based purity 99.99% or higher. Diammonium hydrogen phosphate $((NH_4)_2HPO_4)$ was purchased from EM Science (A.C.S. grade). The volatiles content in $RE_2O_3$ were determined by thermogravimetry using a Netzsch 409 instrument. The stoichiometric mixtures (RE:P = 7:3; total mass about 5 g) were first annealed at 1000 °C for 24 hours. $(NH_4)_2HPO_4$ decomposes at ~190 °C. To prevent $P_2O_5$ loss, the heating rate below 400 °C was selected as 1 K/min, then increased to 5 K/min. The resulting powders were reground, pelletized and annealed at 1500 °C for 48 h (Sm) or 1600 °C for 24 h (La, Pr, Nd, Eu, Gd, Tb). $Pr_7P_3O_{18}$ and $Tb_7P_3O_{18}$ were prepared separately at the University of Bonn, using independently calibrated reagents, $Pr_7P_3O_{18}$ was characterized together with the rest of the samples prepared at Arizona State University.

Powder X-ray diffraction data were collected with a Bruker D2 Phaser diffractometer equipped with a copper tube (Cu-K$\alpha$=1.54051 Å), a 3 mm anti-scattering plate, 2.5° Soller slits, and a Lynxeye detector. NIST Si640c (a = 5.4312 Å) [21] was used as an internal standard for Rietvield refinement with GSAS II software [22]. The space group 14 predicted by DFT for $RE_7P_3O_{18}$ compounds has multiple settings. The CIF files from VASP were converted to standard setting in VESTA software [23] and used for the refinement. Background was initially fitted using fixed points and then refined. Atomic coordinates were fixed to the values obtained by DFT relaxation and were not refined. Atomic displacement parameters, not provided by DFT, were refined isotropically with equivalence constraints applied to each atom type in the structure. The lattice parameters for $Tb_7P_3O_{18}$ were refined using CellRef [24] after visual read-out of the diffraction angles of 66 reflections from an XRPD pattern (Stoe StadiP, cobalt tube, Johannsen monochromator Ge (111), Co-K$\alpha_1$ =1.78901 Å).

Fourier-Transform infrared spectroscopy with attenuated total reflectance (FTIR-ATR) measurements were carried out on a Bruker Platinum ATR spectrometer with an excitation laser source of 532 nm. The data were collected in the transmittance mode and given in cm$^{-1}$. Differential scanning calorimetry measurements on synthesized samples were performed with Netzsch 449 instrument in argon in Pt crucibles with a heating rate 25 K/min.

## 3. Results and discussion

**3.1 Structure and energetics of $RE_7P_3O_{18}$ rare earth oxide-phosphates**

In our ab initio calculations, we used two reported structures of $La_7V_3O_{18}$ [10] as starting points, substituting V with P to generate initial models. The $La_7V_3O_{18}$ structures include the α-phase with $P2_1$ symmetry, known as the low-temperature phase, and the β-phase with $P2_1/c$ symmetry, identified as the high-temperature phase [10]. After full relaxation of various $RE_7P_3O_{18}$ compounds based on these two structures, we found that all structures initially possessing $P2_1$ symmetry transformed into structures with $P2_1/c$ symmetry, while those with $P2_1/c$ symmetry retained their original symmetry. A substantial difference exists between the DFT-optimized structure of $Eu_7O_6(PO_4)_3$ and the structural model previously reported [2], even though both models share the same space group. When optimized using the VASP package, the previously suggested $Eu_7O_6(PO_4)_3$ model [2] exhibits a significantly higher energy compared to the DFT-optimized structure started from the $La_7V_3O_{18}$ prototype. In contrast to the $RE_7P_3O_{18}$ system, our DFT calculations show that all $RE_7V_3O_{18}$ compounds with *$P2_1$* symmetry have lower energies than those with centrosymmetric structures in *$P2_1/c$* [25]. The *$P2_1/c$* structures of $RE_7V_3O_{18}$ remain stable and retain their symmetry after full relaxation. Table 1 and Fig. S1 summarize the formation energies and optimized lattice parameters. The resulting CIF files for all $RE_7P_3O_{18}$ are included in *Supplementary material*.

The crystal structures of β-$La_7O_6(VO_4)_3$ and $La_7O_6(PO_4)_3$ are compared in Fig. 2. The close structural similarity between the two crystal structures (isotypes) with tetrahedral $XO_4^{3-}$ (X: V, P) units and the polycations $[La_7O_6]^{9+}$ is immediately visible. On closer inspection one recognizes the less dense packing of the phosphate anions, in contrast to the vanadate groups, within the polycationic network. The experimentally determined (Rietveld) atomic coordinates for β-$La_7O_6(VO_4)_3$ [10] lead to a rather wide range of distances $d$(V-O) (1.57 to 1.84 Å) and $d$(O-La) (2.12 to 2.71 Å) within the $[OLa_4]$ tetrahedra of the polycation. In contrast, the phosphate tetrahedra in all DFT-optimized oxide-phosphate structures $RE_7O_6(PO_4)_3$ (RE = La – Lu, Y) are far more regular (1.53 ≤ $d$(P-O) ≤ 1.58 Å), as are the distances $d$(O-RE). For $La_7O_6(PO_4)_3$ the range 2.28 ≤ $d$(O-La) ≤ 2.62 Å is found in the relaxed structure. While for the oxide-vanadate the experimentally determined interatomic distances are within chemically acceptable margins, yet with a rather wide spread, the distances in the DFT-relaxed structure of $La_7O_6(PO_4)_3$ fit perfectly into chemical expectation with no unusual spread. To explore this point further we have computed bond valence sums (BVS [26,27]) for β-$La_7O_6(VO_4)_3$ and $La_7O_6(PO_4)_3$. The results are compared in Fig. 3. The radial distribution functions for β-$La_7O_6(VO_4)_3$ and $La_7O_6(PO_4)_3$ are plotted in Fig. 4.

**Table 1.** Calculated formation energies [from Eqs. (1) and (2)] and optimized lattice parameters for $RE_7P_3O_{18}$ compounds. $T_{crit}$ denotes the calculated critical temperatures above which the formation free energies for the reaction $RE_7P_3O_{18} \leftrightarrow 2RE_2O_3 + 3REPO_4$ become negative. The '–' entries in the '$T_{crit}$' column indicate estimated critical temperatures exceeding 2400 K. For comparison, mean melting temperatures estimated from machine learning [28] are also listed.

| $RE_7P_3O_{18}$ | SG | $a$ (Å) | $b$ (Å) | $c$ (Å) | $\beta$ (°) | V/Z (Å³) | $\Delta E_f(1)$ (eV/atom) | $\Delta E_f(2)$ (eV/atom) | $T_{crit}$ $\Delta G_f<0$ (K) | Tm(ML) (K) |
|---|---|---|---|---|---|---|---|---|---|---|
| $La_7P_3O_{18}$ | $P2_1/c$ | 7.07 | 18.74 | 13.64 | 110.0 | 424.5 | 0.013 | -0.011 | 0 | 2030 |
| $Ce_7P_3O_{18}$ | $P2_1/c$ | 7.10 | 18.62 | 13.64 | 109.8 | 424.6 | 0.022 | -0.002 | 0 | 2060 |
| $Pr_7P_3O_{18}$ | $P2_1/c$ | 7.04 | 18.48 | 13.52 | 109.9 | 414.0 | 0.023 | 0.000 | 0 | 1980 |
| $Nd_7P_3O_{18}$ | $P2_1/c$ | 6.99 | 18.38 | 13.42 | 109.7 | 405.6 | 0.024 | 0.001 | 827 | 2050 |
| $Sm_7P_3O_{18}$ | $P2_1/c$ | 6.89 | 18.23 | 13.23 | 109.6 | 391.6 | 0.029 | 0.013 | 1128 | 2300 |
| $Eu_7P_3O_{18}$ | $P2_1/c$ | 6.85 | 18.14 | 13.16 | 109.5 | 385.1 | 0.031 | 0.019 | 1662 | 2080 |
| $Gd_7P_3O_{18}$ | $P2_1/c$ | 6.80 | 18.10 | 13.07 | 109.5 | 379.2 | 0.033 | 0.025 | 1943 | 2120 |
| $Tb_7P_3O_{18}$ | $P2_1/c$ | 6.76 | 18.07 | 13.02 | 109.5 | 374.7 | 0.035 | 0.030 | 2249 | 2390 |
| $Dy_7P_3O_{18}$ | $P2_1/c$ | 6.71 | 18.06 | 12.98 | 109.5 | 370.7 | 0.036 | 0.035 | 2564 | 2380 |
| $Y_7P_3O_{18}$ | $P2_1/c$ | 6.67 | 18.21 | 13.03 | 109.9 | 372.2 | 0.033 | 0.035 | - | 2410 |
| $Ho_7P_3O_{18}$ | $P2_1/c$ | 6.65 | 18.11 | 12.96 | 109.7 | 367.6 | 0.036 | 0.039 | - | 2340 |
| $Er_7P_3O_{18}$ | $P2_1/c$ | 6.60 | 18.09 | 12.93 | 109.7 | 363.5 | 0.037 | 0.044 | - | 2320 |
| $Tm_7P_3O_{18}$ | $P2_1/c$ | 6.56 | 18.05 | 12.88 | 109.8 | 359.0 | 0.037 | 0.048 | - | 2370 |
| $Yb_7P_3O_{18}$ | $P2_1/c$ | 6.52 | 18.00 | 12.85 | 109.8 | 354.8 | 0.037 | 0.051 | - | 2330 |
| $Lu_7P_3O_{18}$ | $P2_1/c$ | 6.49 | 17.98 | 12.77 | 109.8 | 351.0 | 0.037 | 0.055 | - | 2390 |

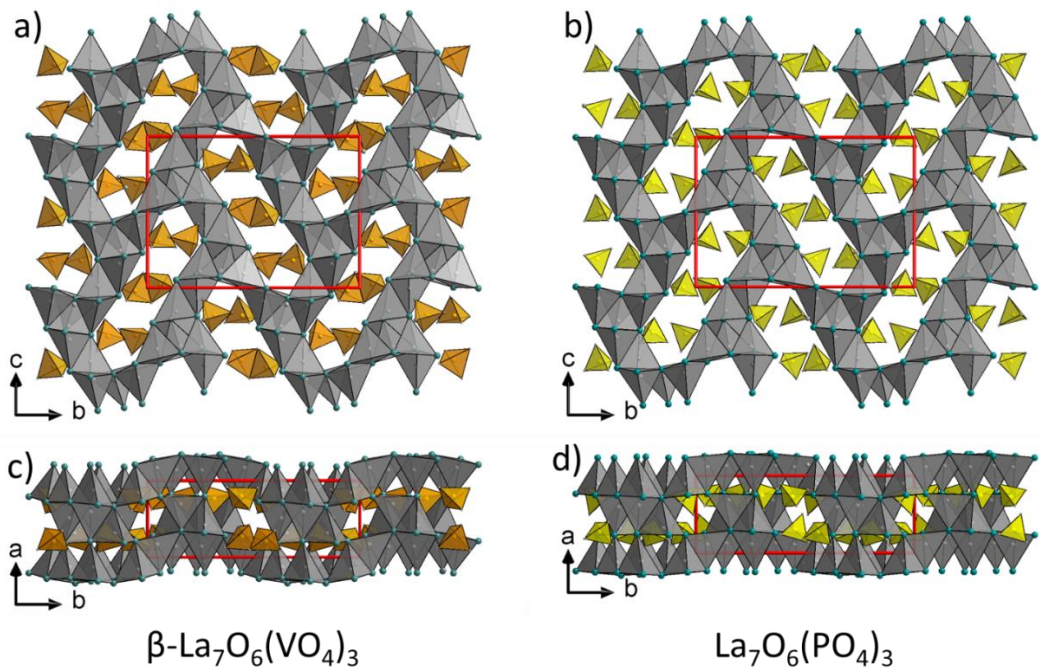

**Fig. 2.** Polyhedral representation of the crystal structures of β-$La_7O_6(VO_4)_3$ (a, c) [10] and $La_7O_6(PO_4)_3$ [this study] (b, d) with $VO_4^{3-}$ tetrahedra (orange), $PO_4^{3-}$ tetrahedra (yellow) and the polycations $[La_7O_6]^{9+}$ (grey with teal $La^{3+}$ ions).

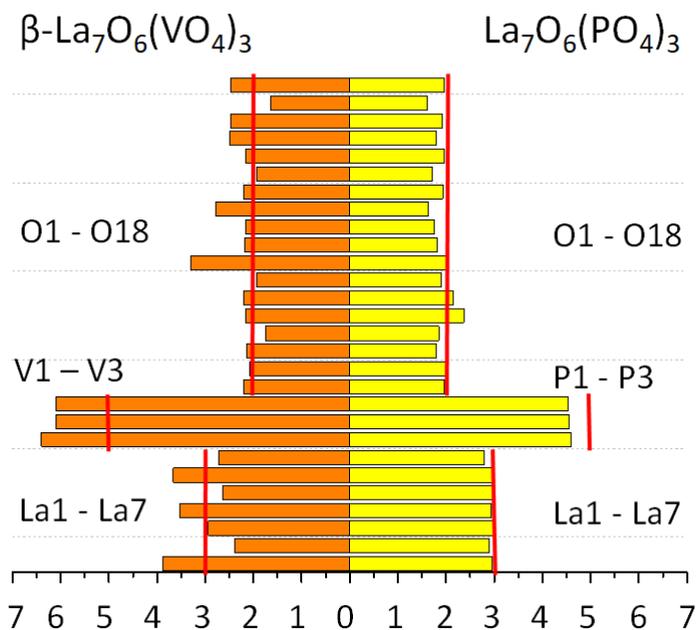

**Fig. 3.** The computed (DIST routine in JANA 2006 [29]) bond valence sums for β-$La_7O_6(VO_4)_3$ and $La_7O_6(PO_4)_3$ (BVS parameters from [30]).

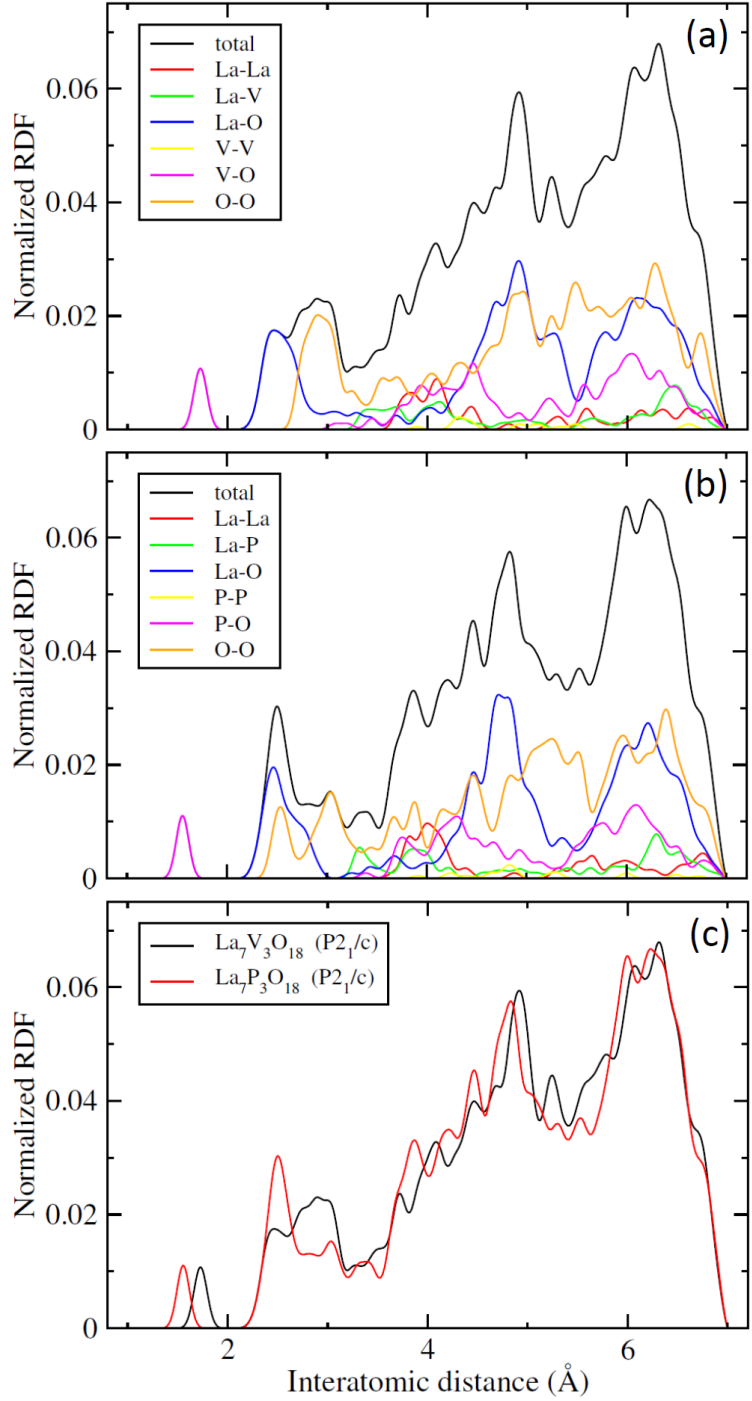

**Fig. 4.** Total and partial radial distribution functions for β-La$_7$O$_6$(VO$_4$)$_3$ (a) and La$_7$O$_6$(PO$_4$)$_3$ (b). (c) Comparison of the total radial distribution functions for the two systems.

The formation energies of $RE_7P_3O_{18}$ compounds from $RE_3PO_7$ and $REPO_4$, or from $RE_2O_3$ and $REPO_4$, were calculated according to the definitions in Eqs. (1) and (2). As shown in Table 1, $La_7P_3O_{18}$ and $Ce_7P_3O_{18}$ exhibit negative formation energies according to Eq. (2), indicating that their formation from the corresponding sesquioxides ($La_2O_3$ and $Ce_2O_3$) and orthophosphates ($LaPO_4$ and $CePO_4$) is exothermic. However, for La and Ce, $RE_7P_3O_{18}$ is unstable at 0 K with respect to decomposition into $RE_3PO_7$ and $REPO_4$. For all other $RE_7P_3O_{18}$ compounds, the formation energies are positive by both Eq. (1) and Eq. (2), suggesting that if these compounds are synthesized at high temperatures they must be stabilized by entropy effects. Based on convex hull analysis, $RE_7P_3O_{18}$ compounds with RE = La to Dy are predicted to decompose into $RE_3PO_7$ and $REPO_4$ (i.e., via the decomposition reaction $RE_7P_3O_{18} \rightarrow 2RE_3PO_7 + REPO_4$). For the heavier rare earth elements, from Ho to Lu (including Y), the corresponding $RE_7P_3O_{18}$ compounds are expected to decompose into $RE_2O_3$ and $REPO_4$ (i.e., via the decomposition reaction $RE_7P_3O_{18} \rightarrow 2RE_2O_3 + 3REPO_4$), since the $RE_3PO_7$ phase is unstable for these elements [7]. As shown in Table 1, all $RE_7P_3O_{18}$ compounds lie at most 55 meV/atom above the convex hull.

**3.2 $RE_7P_3O_{18}$ compounds stabilized by entropy effects**

As discussed above, $RE_7P_3O_{18}$ compounds are unstable at 0 K. Entropy contributions play a crucial role in stabilizing phases at elevated temperatures [19]. Using the method described in Section 2, we therefore evaluated the entropy contributions to phase stability to determine the critical temperatures above which $RE_7P_3O_{18}$ compounds can be stabilized. Predicting these critical temperatures is particularly valuable, as it provides theoretical guidance for the experimental synthesis of $RE_7P_3O_{18}$. In our experimental work, rare earth oxides were typically employed as the metal source, consistent with the syntheses of $Eu_7P_3O_{18}$ and $Sm_7P_3O_{18}$ reported by Glaum's group at the University of Bonn [2]. Accordingly, the formation free energy calculations were based on the reaction $RE_7P_3O_{18} \leftrightarrow 2RE_2O_3 + 3REPO_4$.

To evaluate the entropy contributions of bulk $RE_7P_3O_{18}$, $RE_2O_3$, and $REPO_4$, ab initio molecular dynamics (AIMD) simulations were performed at 1500 K for typically more than 24,000 steps using the SLUSCHI package [31,32] within the isothermal–isobaric (*NPT*) ensemble. Entropy data were extracted by discarding the initial 4,000, 8,000, and 12,000 steps, and the results from these three sampling windows were averaged. Using these data, the formation free energies of $RE_7P_3O_{18}$ as a function of temperature were calculated. From this analysis, the critical temperatures ($T_{crit}$) - defined as the temperatures above which the $RE_7P_3O_{18}$ phases become thermodynamically stable due to entropy contributions, i.e., where the formation free energy $\Delta G_f$ becomes negative ($\Delta G_f < 0$) - were determined. In these calculations, the total energies in Eq. (2) were replaced by the corresponding free energies. As reported in Ref. [20], vibrational entropy provides the dominant contribution to the total entropy, while configurational entropy is typically an order of magnitude smaller in liquids and negligible in solids. Therefore, only vibrational entropy contributions were included in the free energy calculations.

The calculated critical temperature results are listed in Table 1 and plotted in Fig. 5. The calculated thermodynamic properties are also presented in *Supplementary material* (Table S1). From Table 1 and Fig. 5, we observe that $RE_7P_3O_{18}$ compounds with RE = La–Tb can be formed from $RE_2O_3$ and $REPO_4$ at temperatures below their estimated melting points. In contrast, for $RE_7P_3O_{18}$ compounds with RE = Dy–Lu (including Y), the critical temperatures exceed their melting points, suggesting that these phases cannot be stabilized prior to melting. These predictions are in good agreement with the available experimental observations, as shown in Fig. 1.

As shown in Fig. 5, the critical temperature exhibits an approximately linear correlation with the ionic radius of the RE cation for $RE_7P_3O_{18}$ compounds with RE = Nd–Dy. Similarly, the formation energies defined by Eq. (2) also display a linear dependence on the RE ionic radius [Fig. 5(b)]. These results indicate that geometric factors, particularly variations in ionic size, play a dominant role in governing both the thermodynamic stability of the compounds and their critical stabilization temperatures. This trend is expected, as the electronic structures of the RE elements are largely similar, while their ionic radii vary substantially across the series.

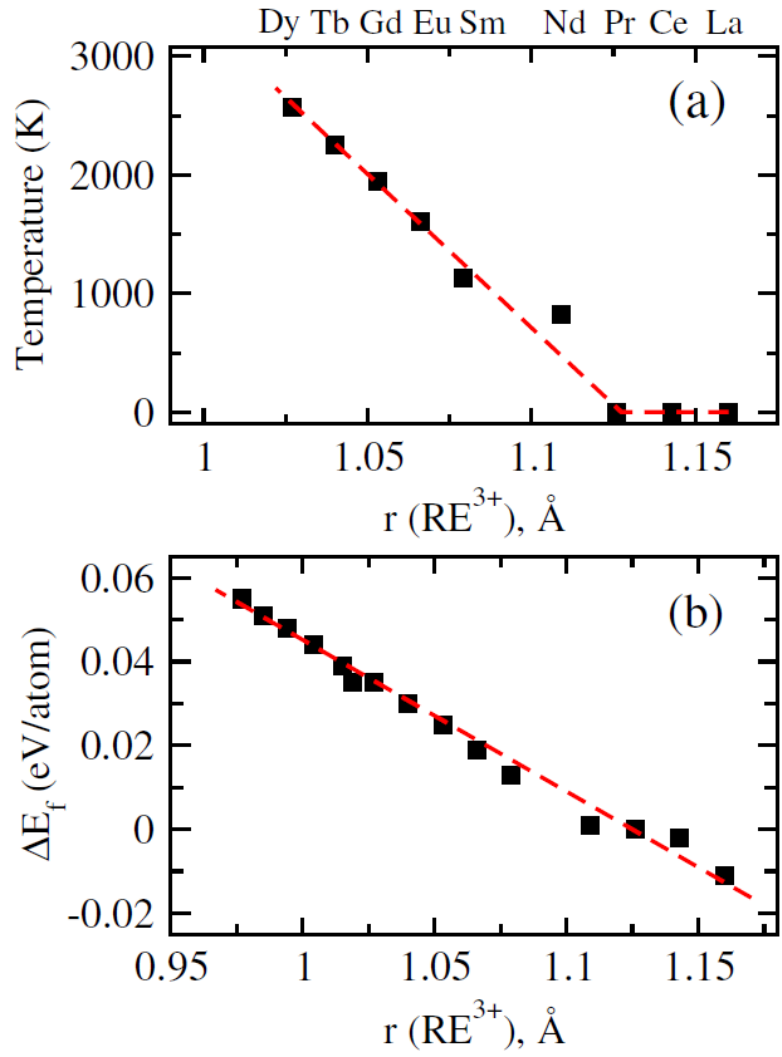

**Fig. 5.** Theoretical critical stabilization temperature (a) and formation energy (b) as functions of the ionic radius of the rare-earth element [33].

### 3.3. Synthesis and characterization of oxide-phosphates

In this work we synthesized $RE_7P_3O_{18}$ phases from rare earth sesquioxides $RE_2O_3$ and diammonium hydrogen phosphate $((NH_4)_2HPO_4)$. This reaction can proceed through formation of different metastable phases, such as rare earth polyphosphates. However, $RE_7P_3O_{18}$ were also previously synthesized by direct reaction of rare earth orthophosphates $REPO_4$ with $RE_2O_3$ [2,6] which was studied computationally in this work. The FTIR spectra (Fig.6) of the synthesized samples show almost identical patterns of bending and stretching O-P vibrations between 100 $cm^1$ and 1100 $cm^{-1}$, which are distinctly different form previously reported for $RE_3PO_7$ [34].

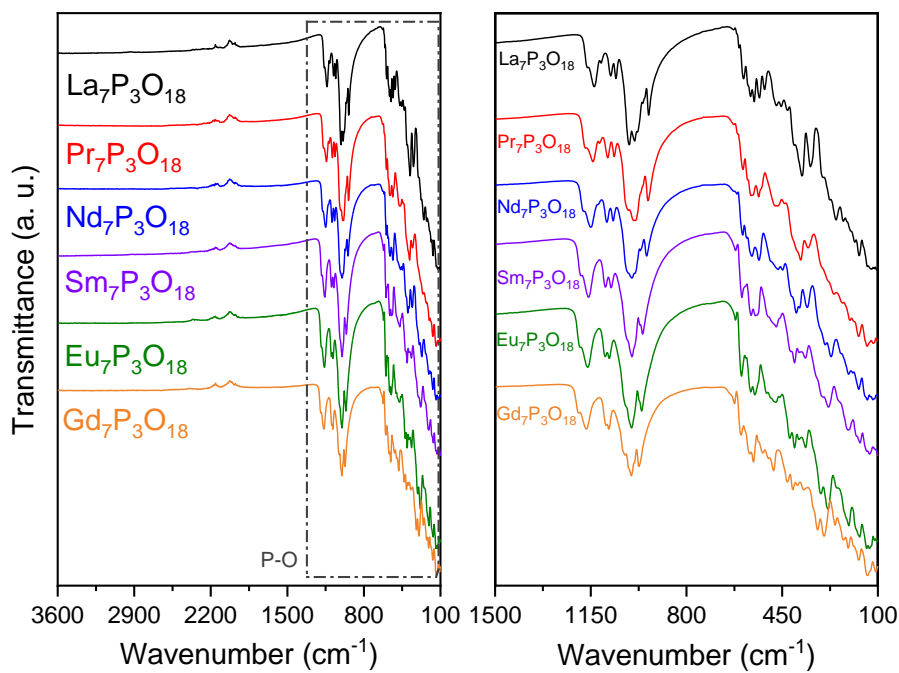

**Fig. 6.** FTIR–ATR spectra of $Ln_7P_3O_{18}$ (Ln = La, Pr, Nd, Sm, Eu, Gd), showing characteristic P-O vibrations in the 1400–100 $cm^{-1}$ (magnified on the right) and a weak $CO_2$ background near 2100 $cm^{-1}$.

Differential scanning calorimetry measurements performed for all samples from room temperature to synthesis temperature 1400 °C and did not indicate any pronounced peaks in a heat flow trace on heating or on cooling, indicating absence of reversible phase transformations.

The structures predicted from DFT were used for Rietveld refinement of synthesized $RE_7P_3O_{18}$ samples (Table 2). To validate accuracy of predicted structures, the atomic positions were not refined. Notably, the quality of refinement, as indicated by weighted residual factor (Rwp), increased from $La_7P_3O_{18}$ (Rwp 10.8 %) to $Gd_7P_3O_{18}$ (Rwp 2.2 %). On refinement of the cell parameters, the change in *a* and *c* parameters compared with computed values was insignificant, however refined *b* parameter was found to be smaller than predicted. Consistent with refinement metrics, the largest difference in *b* from computed value was observed for $La_7P_3O_{18}$ (-3.7 %) and the smallest for $Gd_7P_3O_{18}$ (-1.5 %). The difference in refined beta angle was also largest for $La_7P_3O_{18}$ (-0.9 %) and smallest for $Gd_7P_3O_{18}$ (-0.3 %). The orthophosphate phase $REPO_4$ (monazite) in the amounts varying from 4.6 to 6.7 wt% was detected in all samples from solid state synthesis aimed for 7:3 RE/P stoichiometry and analyzed by Rietveld refinement. No apparent extra reflections were detected in the patterns (Fig. 7).

The synthesis of $Ce_7P_3O_{18}$ was previously achieved by laser melting of $CePO_4$ in 5 vol% $H_2$ balanced with Ar (Supplementary Information in Ref. [7]). In that work, in the absence of structural model, $Ce_7P_3O_{18}$ XRD pattern was indexed using $Nd_7P_3O_{18}$ and $La_7P_3O_{18}$ entries in PDF database, with reference to Serra et al. [6]. Rietveld refinement of the same XRD pattern of $Ce_7P_3O_{18}$ from laser melting has since been carried out using a DFT-derived structure (Fig S1 in Supplementary Information). Notably, the refined cell parameters for $Ce_7P_3O_{18}$ are consistent with the trend obtained on refinement of $RE_7P_3O_{18}$ samples which were synthesized in this work and did not undergo melting.

**Table 2.** The results of Rietveld refinement of cell parameters of synthesized $RE_7P_3O_{18}$ samples using DFT derived structure. Si 640c was used as internal standard for refinement of all XRD patterns, except $Ce_7P_3O_{18}$ and $Tb_7P_3O_{18}$.

| RE | a (Å) | b (Å) | c (Å) | V (Å$^3$) | β (°) | GoF | $R_{wp}$, % | $REPO_4$ wt% |
|---|---|---|---|---|---|---|---|---|
| La | 7.0863(5) | 18.0757(4) | 13.674(2) | 1634.94(6) | 111.022(2) | 4.87 | 10.82 | 4.6(1) |
| Ce[a)] | 7.021(1) | 17.937(1) | 13.550(3) | 1594.0(2) | 110.901(3) | 1.23 | 5.33 | -- |
| Pr | 6.9694(4) | 17.8568(3) | 13.472(1) | 1566.66(4) | 110.865(2) | 4.10 | 7.58 | 6.4(2) |
| Nd | 6.9328(5) | 17.8416(3) | 13.378(2) | 1548.30(6) | 110.666(2) | 3.10 | 5.96 | 6.6(2) |
| Sm | 6.8531(3) | 17.8242(3) | 13.198(1) | 1512.18(4) | 110.288(1) | 2.02 | 2.99 | 4.8(1) |
| Eu | 6.8150(4) | 17.8224(3) | 13.126(1) | 1496.68(5) | 110.148(2) | 2.05 | 2.73 | 6.7(2) |
| Gd | 6.7926(5) | 17.8382(4) | 13.042(2) | 1486.14(6) | 109.882(3) | 1.78 | 2.22 | 6.2(2) |
| Tb[b)] | 6.734(8) | 17.713(7) | 12.943(2) | 1451.6(1) | 109.90(1) | --- | --- | --- |

[a)] $Ce_7P_3O_{18}$ sample prepared by laser melting of $CePO_4$ [7], contained residual $CePO_4$ and $Ce_3PO_7$ phases.
[b)] $Tb_7P_3O_{18}$ lattice parameters determined with 66 reflections from visual read-out of an XRPD pattern.

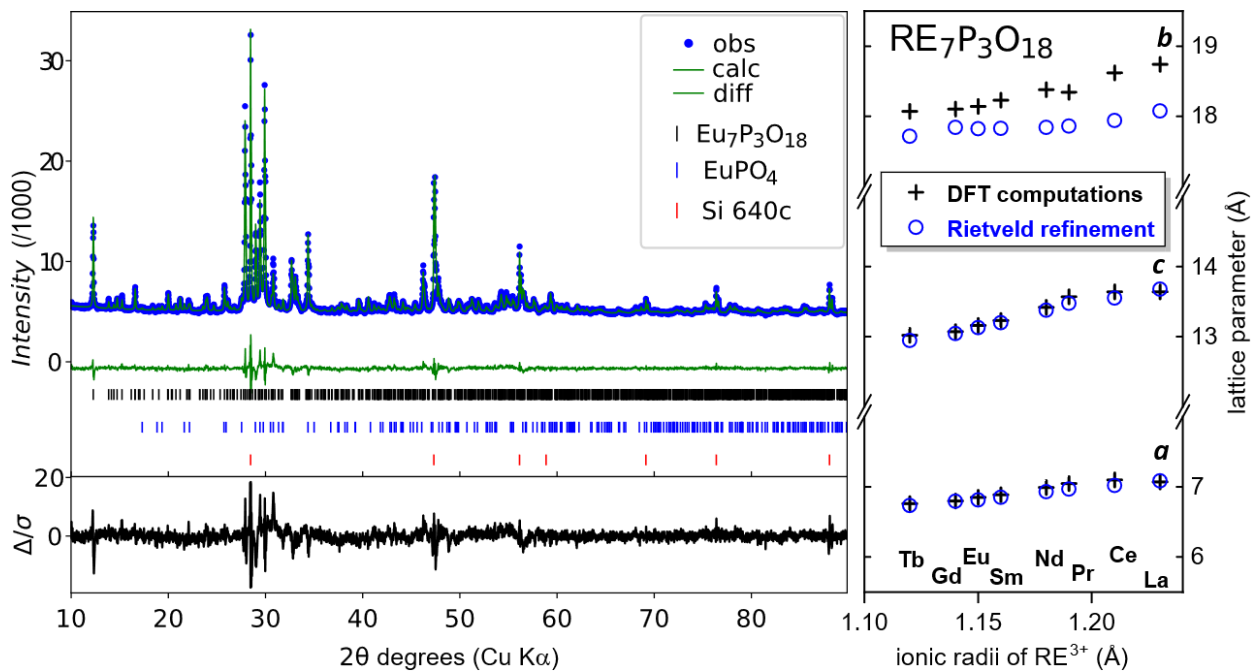

**Fig. 7.** Left: Rietveld refinement of $Eu_7P_3O_{18}$ sample using DFT derived structure. Si 640c added as internal standard for cell parameters refinement in all samples except Tb and Ce compounds. Atomic positions were not refined. Refinement metrics: Rwp 2.73 % GoF 2.05. Right: Unit cell parameters for synthesized $RE_7P_3O_{18}$ (RE = La, Ce, Pr, Nd, Sm, Eu, Gd) from Rietveld refinement compared with computed values. The symbol size for experimental values exceeds error bars.

## 4. Summary and Conclusions

The rare earth oxide-phosphates with stoichiometry $RE_7P_3O_{18}$ were first reported more than half a century ago. However, their structure has not been solved until now due to difficulty of growing single crystals and difficulty of structure solution from powder diffraction data given their low symmetry and complex structure. This has hampered their potential applications for development of the new refractory, magnetic and catalyst materials. Since rare earth orthophosphates ($REPO_4$) and rare earth orthovanadates ($REVO_4$) are known to adapt the same structure types across rare earth series, we used recently reported structure for $La_7V_3O_{18}$ to computationally predict structure and stability for $RE_7P_3O_{18}$ for all lanthanides and yttrium. The $RE_7P_3O_{18}$ compounds were found to be thermodynamically unstable at 0 K, tending to decompose into $REPO_4$ and $RE_3PO_7$ or $RE_2O_3$. However, for RE = La–Tb, they may be stabilized

18by entropy contributions, consistent with the observed synthesis behavior. The predicted structures were successfully used for Rietveld analyses of X-ray diffraction patterns. The analyses revealed the presence of 4-6 wt % of REPO$_4$ in independently synthesized samples prepared in RE$_7$P$_3$O$_{18}$ stoichiometry. This may indicate that the real structure deviates from the 7:3 RE/P ratio.

Acknowledgments
**Acknowledgments**

This research was supported by US Department of Defense Army Research Office Award number W911NF-23-2-0145, with use of Research Computing at Arizona State University.

*Supplementary material for the manuscript*

# Structure and stability of 7:3 rare earth oxyphosphates:

# a combined ab initio and experimental study

Ligen Wang[a], Konrad Burkmann[b], Sergey V. Ushakov[b], Edric X. Wang[a,c], Jared Matteucci[b], Mara Scheuermann[d], Erik Melnitschuk[d], Robert Glaum[d], Hongwu Xu[b], Elizabeth J. Opila[e], Alexandra Navrotsky[a,b], Qi-Jun Hong[a],*

[a] Materials Science and Engineering, Arizona State University, Tempe, AZ 85287, USA
[b] Navrotsky-Eyring Center for Materials of the Universe, School of Molecular Sciences, Arizona State University, Tempe, AZ 85287, USA
[c] Department of Materials Science and Engineering, University of Illinois at Urbana-Champaign, Urbana, IL 61801, USA
[d] Department of Inorganic Chemistry, Rheinische Friedrich-Wilhelms-Universität Bonn, Gerhard-Domagk-Straße 1, 53121 Bonn, Germany
[e] Department of Materials Science and Engineering, University of Virginia, Charlottesville, VA 22903, USA

**Figure S1.** The computed cell parameters and volumes of $RE_7P_3O_{18}$ for RE = La–Lu (excluding Pm) and Y.

**Figure S2.** Rietvield refinement of XRD diffraction pattern (CuKa) of $Ce_7P_3O_{18}$ sample prepared by laser melting in earlier reported study (Supplementary Information in Ref.[1]). The DFT derived structure model for $Ce_7P_3O_{18}$ was used without refinement of the atomic positions. Refinement metrics: Rwp 5.33 %, GoF 1.23.

**Table S1.** Thermodynamic properties and critical temperature ($T_{crit}$) above which the formation free energies for the reaction $RE_7P_3O_{18} \leftrightarrow 2RE_2O_3 + 3REPO_4$ become negative.

CIF files from DFT derived structures for $RE_7P_3O_{18}$ (RE = La-Lu and Y)

CCDC 2492364 - 2492378 contain the supplementary crystallographic data for this paper. These data can be obtained free of charge via www.ccdc.cam.ac.uk/data_request/cif, or by emailing data_request@ccdc.cam.ac.uk, or by contacting The Cambridge Crystallographic Data Centre, 12 Union Road, Cambridge CB2 1EZ, UK; fax: +44 1223 336033.

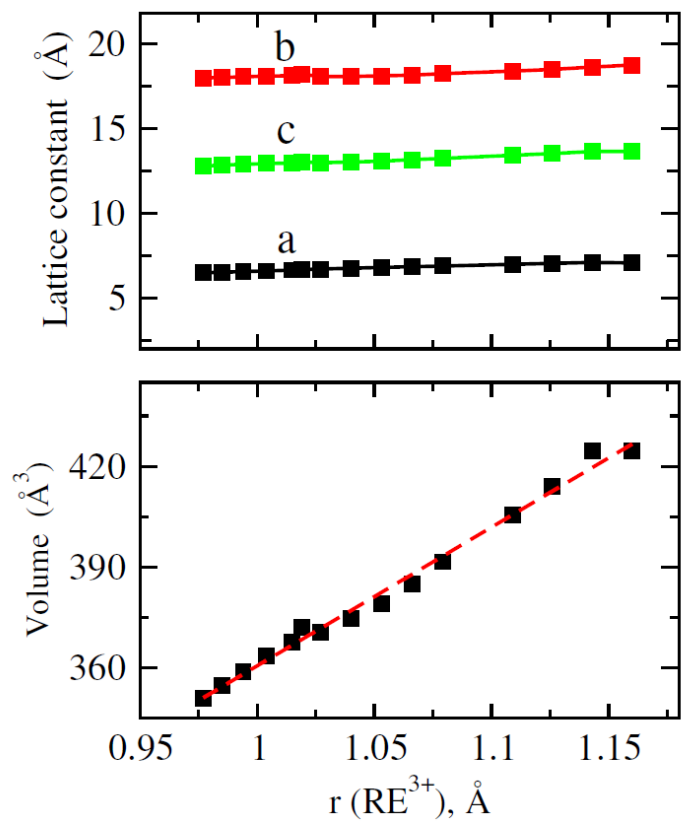

**Figure S1.** The computed cell parameters and volumes of RE$_7$P$_3$O$_{18}$ for RE = La–Lu (excluding Pm) and Y.

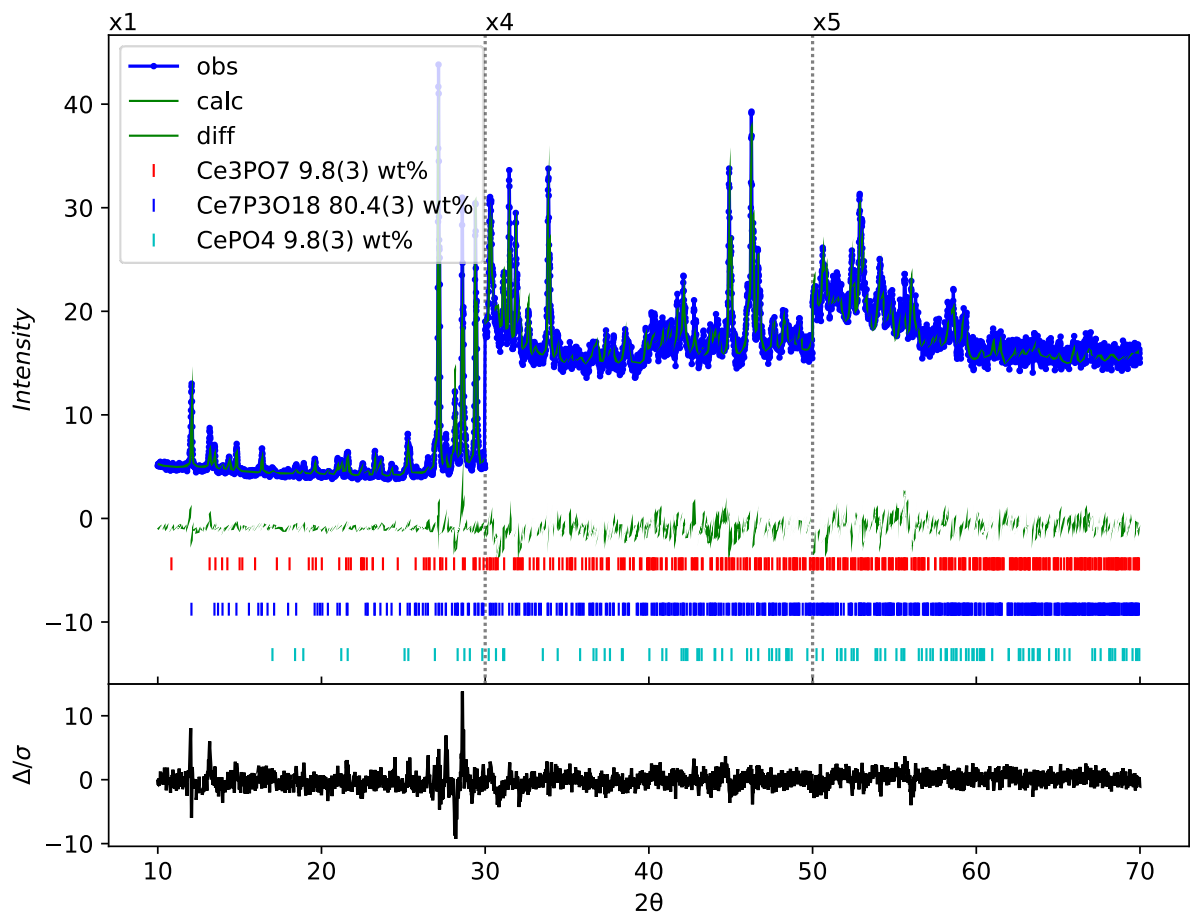

**Figure S2.** Rietvield refinement of XRD diffraction pattern (CuKa) of $Ce_7P_3O_{18}$ sample prepared by laser melting in earlier reported study (Supplementary Information in Ref. [1]). The DFT derived structure model for $Ce_7P_3O_{18}$ was used without refinement of the atomic positions. Refinement metrics: Rwp 5.33 %, GoF 1.23.

**Table S1.** Thermodynamic properties and critical temperature ($T_{crit}$) above which the formation free energies for the reaction $RE_7P_3O_{18} \leftrightarrow 2RE_2O_3 + 3REPO_4$ become negative.

| System | SG | T(MD) [K] | H [EV/atom] | S [J/K/mol atom] | $\Delta H_f$ [J/mol atom] | $\Delta S_f$ [J/K/mol atom] | $T_{crit}$ [K] |
|---|---|---|---|---|---|---|---|
| Nd$_7$P$_3$O$_{18}$ | P2$_1$/c | 1500 | -7.862 | 59.188 | 900 | 1.088 | 827 |
| Nd$_2$O$_3$ | Ia-3 | 1500 | -8.030 | 64.663 | | | |
| NdPO$_4$ | P2$_1$/c | 1500 | -7.783 | 54.454 | | | |
| Sm$_7$P$_3$O$_{18}$ | P2$_1$/c | 1500 | -7.877 | 58.568 | 1776 | 1.575 | 1128 |
| Sm$_2$O$_3$ | Ia-3 | 1500 | -8.082 | 64.541 | | | |
| SmPO$_4$ | I4$_1$/amd | 1500 | -7.792 | 52.801 | | | |
| Eu$_7$P$_3$O$_{18}$ | P2$_1$/c | 1500 | -7.896 | 57.363 | 2436 | 1.465 | 1662 |
| Eu$_2$O$_3$ | Ia-3 | 1500 | -8.130 | 62.792 | | | |
| EuPO$_4$ | I4$_1$/amd | 1500 | -7.805 | 52.068 | | | |
| Gd$_7$P$_3$O$_{18}$ | P2$_1$/c | 1500 | -7.900 | 58.475 | 3412 | 1.756 | 1943 |
| Gd$_2$O$_3$ | Ia-3 | 1500 | -8.157 | 63.919 | | | |
| GdPO$_4$ | I4$_1$/amd | 1500 | -7.813 | 52.718 | | | |
| Tb$_7$P$_3$O$_{18}$ | P2$_1$/c | 1500 | -7.904 | 58.190 | 3806 | 1.692 | 2249 |
| Tb$_2$O$_3$ | Ia-3 | 1500 | -8.175 | 63.315 | | | |
| TbPO$_4$ | I4$_1$/amd | 1500 | -7.814 | 52.711 | | | |
| Dy$_7$P$_3$O$_{18}$ | P2$_1$/c | 1500 | -7.905 | 58.162 | 4199 | 1.638 | 2564 |
| Dy$_2$O$_3$ | Ia-3 | 1500 | -8.191 | 63.574 | | | |
| DyPO$_4$ | I4$_1$/amd | 1500 | -7.814 | 52.608 | | | |